\documentclass[twocolumn,showpacs,aps,prl,superscriptaddress]{revtex4}

\usepackage{graphicx}
\usepackage{dcolumn}
\usepackage{amsmath}
\usepackage{epsfig}
\usepackage{multirow}

\RequirePackage{xspace}

\usepackage{relsize}
\def\babar{\mbox{\slshape B\kern-0.1em{\smaller A}\kern-0.1em
    B\kern-0.1em{\smaller A\kern-0.2em R}}}

\def\epem       {\ensuremath{e^+e^-}\xspace}

\def\gaga  {\ensuremath{\gamma\gamma}\xspace}  

\def\qqbar {\ensuremath{q\overline q}\xspace}
\def\u     {\ensuremath{u}\xspace}
\def\ubar  {\ensuremath{\overline u}\xspace}

\def\dbar  {\ensuremath{\overline d}\xspace}

\def\s     {\ensuremath{s}\xspace}

\def\c     {\ensuremath{c}\xspace}
\def\cbar  {\ensuremath{\overline c}\xspace}
\def\ccbar {\ensuremath{c\overline c}\xspace}
\def\b     {\ensuremath{b}\xspace}
\def\bbar  {\ensuremath{\overline b}\xspace}

\def\piz   {\ensuremath{\pi^0}\xspace}

\def\pip   {\ensuremath{\pi^+}\xspace}
\def\pim   {\ensuremath{\pi^-}\xspace}

\def\pipm  {\ensuremath{\pi^\pm}\xspace}

\def\Kbar  {\kern 0.2em\overline{\kern -0.2em K}{}\xspace}

\def\Kz    {\ensuremath{K^0}\xspace}
\def\Kzb   {\ensuremath{\Kbar^0}\xspace}
\def\KzKzb {\ensuremath{\Kz \kern -0.16em \Kzb}\xspace}
\def\Kp    {\ensuremath{K^+}\xspace}
\def\Km    {\ensuremath{K^-}\xspace}

\def\KpKm  {\ensuremath{\Kp \kern -0.16em \Km}\xspace}
\def\KS    {\ensuremath{K^0_{\scriptscriptstyle S}}\xspace}

\def\Kstar   {\ensuremath{K^*}\xspace}

\def\Kstarp  {\ensuremath{K^{*+}}\xspace}
\def\Kstarm  {\ensuremath{K^{*-}}\xspace}
\def\Kstarpm {\ensuremath{K^{*\pm}}\xspace}

\def\Dbar    {\kern 0.2em\overline{\kern -0.2em D}{}\xspace}

\def\Dz      {\ensuremath{D^0}\xspace}
\def\Dzb     {\ensuremath{\Dbar^0}\xspace}
\def\DzDzb   {\ensuremath{\Dz {\kern -0.16em \Dzb}}\xspace}
\def\Dp      {\ensuremath{D^+}\xspace}
\def\Dm      {\ensuremath{D^-}\xspace}

\def\DpDm    {\ensuremath{\Dp {\kern -0.16em \Dm}}\xspace}
\def\Dstar   {\ensuremath{D^*}\xspace}

\def\Dstarp  {\ensuremath{D^{*+}}\xspace}
\def\Dstarm  {\ensuremath{D^{*-}}\xspace}

\def\B       {\ensuremath{B}\xspace}
\def\Bbar    {\kern 0.18em\overline{\kern -0.18em B}{}\xspace}
\def\Bb      {\ensuremath{\Bbar}\xspace}
\def\BB      {\ensuremath{B\Bbar}\xspace} 
\def\Bz      {\ensuremath{B^0}\xspace}
\def\Bzb     {\ensuremath{\Bbar^0}\xspace}
\def\BzBzb   {\ensuremath{\Bz {\kern -0.16em \Bzb}}\xspace}
\def\Bu      {\ensuremath{B^+}\xspace}
\def\Bub     {\ensuremath{B^-}\xspace}

\def\Bm      {\ensuremath{\Bub}\xspace}

\def\BpBm    {\ensuremath{\Bu {\kern -0.16em \Bub}}\xspace}

\def\BorBbar    {\kern 0.18em\optbar{\kern -0.18em B}{}\xspace}
\def\DorDbar    {\kern 0.18em\optbar{\kern -0.18em D}{}\xspace}
\def\KorKbar    {\kern 0.18em\optbar{\kern -0.18em K}{}\xspace}

\mathchardef\Upsilon="7107
\def\Y#1S{\ensuremath{\Upsilon{(#1S)}}\xspace}

\def\FourS {\Y4S}

\mathchardef\Deltares="7101
\mathchardef\Xi="7104
\mathchardef\Lambda="7103
\mathchardef\Sigma="7106
\mathchardef\Omega="710A

\def\Deltabar{\kern 0.25em\overline{\kern -0.25em \Deltares}{}\xspace}
\def\Lbar{\kern 0.2em\overline{\kern -0.2em\Lambda\kern 0.05em}\kern-0.05em{}\xspace}
\def\Sigbar{\kern 0.2em\overline{\kern -0.2em \Sigma}{}\xspace}
\def\Xibar{\kern 0.2em\overline{\kern -0.2em \Xi}{}\xspace}
\def\Obar{\kern 0.2em\overline{\kern -0.2em \Omega}{}\xspace}
\def\Nbar{\kern 0.2em\overline{\kern -0.2em N}{}\xspace}
\def\Xb{\kern 0.2em\overline{\kern -0.2em X}{}\xspace}

\newcommand{\tev}{\ensuremath{\mathrm{\,Te\kern -0.1em V}}\xspace}
\newcommand{\gev}{\ensuremath{\mathrm{\,Ge\kern -0.1em V}}\xspace}
\newcommand{\mev}{\ensuremath{\mathrm{\,Me\kern -0.1em V}}\xspace}
\newcommand{\kev}{\ensuremath{\mathrm{\,ke\kern -0.1em V}}\xspace}
\newcommand{\ev}{\ensuremath{\mathrm{\,e\kern -0.1em V}}\xspace}
\newcommand{\gevc}{\ensuremath{{\mathrm{\,Ge\kern -0.1em V\!/}c}}\xspace}
\newcommand{\mevc}{\ensuremath{{\mathrm{\,Me\kern -0.1em V\!/}c}}\xspace}
\newcommand{\gevcc}{\ensuremath{{\mathrm{\,Ge\kern -0.1em V\!/}c^2}}\xspace}
\newcommand{\mevcc}{\ensuremath{{\mathrm{\,Me\kern -0.1em V\!/}c^2}}\xspace}

\def\mus  {\ensuremath{\rm \,\mus}\xspace}

\def\mus        {\ensuremath{\,\mu{\rm s}}\xspace}    

\def\ra                 {\ensuremath{\rightarrow}\xspace}

\def\pep2{PEP-II}

\def\gsim{{~\raise.15em\hbox{$>$}\kern-.85em
          \lower.35em\hbox{$\sim$}~}\xspace}
\def\lsim{{~\raise.15em\hbox{$<$}\kern-.85em
          \lower.35em\hbox{$\sim$}~}\xspace}

\def\CP                {\ensuremath{C\!P}\xspace}

\xspace

\newcommand{\jplBase}        {Phys.\ Lett.\xspace}

\newcommand{\plb}       [1]  {\jplBase\ B~{\bf #1}}

\def\jetset74   {\mbox{\tt Jetset \hspace{-0.5em}7.\hspace{-0.2em}4}\xspace}

\newcommand{\BABARPubYear}    {07}
\newcommand{\BABARPubNumber}  {047}

\newcommand{\SLACPubNumber} {12727}

\def\figurebox#1#2#3{%
    \def\arg{#3}%
    \ifx\arg\empty
    {\hfill\vbox{\hsize#2\hrule\hbox to #2{\vrule\hfill\vbox to #1{\hsize#2\vfill}\vrule}\hrule}\hfill}%
    \else
    {\hfill\epsfbox{#3}\hfill}%
    \fi}

\newcommand{\DE}{\ensuremath{\Delta E}\xspace}
\newcommand{\mES}{\ensuremath{m_\mathrm{ES}}\xspace}
\newcommand{\mDz}{\ensuremath{m_{\Dz}}\xspace}
\newcommand{\mD}{\ensuremath{m_{D}}\xspace}
\newcommand{\hz}{\ensuremath{h^0}\xspace}

\newcommand{\dm}{\ensuremath{\Delta m}\xspace}
\newcommand{\dt}{\ensuremath{\Delta t}\xspace}
\newcommand{\sigmadt}{\ensuremath{\sigma_{\Delta t}}\xspace}

\newcommand{\A}{\ensuremath{{\cal A}}\xspace}
\newcommand{\Abar}{\ensuremath{\overline{\A}\xspace}}
\newcommand{\fbar}{\ensuremath{\overline{f}}\xspace}
\newcommand{\Af}{\ensuremath{\A_f}\xspace}

\newcommand{\Abarfbar}{\ensuremath{\Abar_{\fbar}}\xspace}
\newcommand{\D}{\ensuremath{D}\xspace}
\newcommand{\DDstar}{\ensuremath{D^{(*)}}\xspace}
\newcommand{\DDstarz}{\ensuremath{D^{(*)0}}\xspace}

\newcommand{\AD}{\ensuremath{\A_{\Dz}}\xspace}
\newcommand{\ADbar}{\ensuremath{\A_{\Dzb}}\xspace}

\newcommand{\etap}{\ensuremath{\eta^{\prime}}\xspace}
\newcommand{\etal}{{\it et al.}}

\newcommand{\Btag}{\ensuremath{B_{\rm{tag}}}\xspace}

\newcommand{\ttag}{\ensuremath{t_{\rm{tag}}}\xspace}
\newcommand{\trec}{\ensuremath{t_{\rm{rec}}}\xspace}

\newcommand{\msp}{\ensuremath{m^{2}_{+}}\xspace}
\newcommand{\msm}{\ensuremath{m^{2}_{-}}\xspace}
\newcommand{\mspm}{\ensuremath{m^{2}_{\pm}}\xspace}

\newcommand{\msqKspm}{\ensuremath{m^{2}_{\KS\pipm}}\xspace}

\newcommand{\sinbb}{\ensuremath{\sin2\beta}\xspace}
\newcommand{\cosbb}{\ensuremath{\cos2\beta}\xspace}
\newcommand{\abslambda}{\ensuremath{|\lambda|}\xspace}

\begin{document}

\begin{flushleft}
\babar-PUB-\BABARPubYear/\BABARPubNumber\\
SLAC-PUB-\SLACPubNumber\\
\end{flushleft}

\begin{flushright}
\end{flushright}

\title{
{\large \bf
Measurement of {\boldmath \cosbb} in {\boldmath $\Bz\ra \DDstar\hz$} Decays
with a Time-Dependent  
Dalitz Plot Analysis of {\boldmath $D\ra\KS\pip\pim$}
} 
}

\author{B.~Aubert}
\author{M.~Bona}
\author{D.~Boutigny}
\author{Y.~Karyotakis}
\author{J.~P.~Lees}
\author{V.~Poireau}
\author{X.~Prudent}
\author{V.~Tisserand}
\author{A.~Zghiche}
\affiliation{Laboratoire de Physique des Particules, IN2P3/CNRS et Universit\'e de Savoie, F-74941 Annecy-Le-Vieux, France }
\author{J.~Garra~Tico}
\author{E.~Grauges}
\affiliation{Universitat de Barcelona, Facultat de Fisica, Departament ECM, E-08028 Barcelona, Spain }
% ...
\collaboration{The \babar\ Collaboration}
\noaffiliation

\date{\today}

\begin{abstract}
We study the time-dependent Dalitz plot of $D \ra\KS\pip\pim$ in
$\Bz\ra \DDstar\hz$ decays, where \hz is a $\piz$, $\eta$, $\etap$,
or $\omega$ meson and $\Dstar\ra D\piz$,
using a data sample of $383\times 10^{6}$
$\FourS\ra\BB$ decays collected with the \babar\ detector.
We determine $\cosbb =  0.42 \pm
0.49 \pm 0.09 \pm 0.13$, $\sinbb =  0.29 \pm 0.34 \pm 0.03 \pm 0.05$, and
$\abslambda = 1.01 \pm 0.08 \pm 0.02 $,
where the first error is
statistical, the second is the experimental systematic uncertainty, and
the third, where given, is the Dalitz model uncertainty.
Assuming the world average value for $\sinbb$ and $\abslambda=1$, $\cosbb>0$ is
preferred over $\cosbb<0$ at 86\% confidence level. 
\end{abstract}


\maketitle


Time-dependent \CP asymmetries in \Bz meson decays, resulting from the
interference between decays with and without \Bz-\Bzb mixing, have been studied
with high precision in $\b\ra\ccbar \s $ decay modes by the \babar\ and
Belle collaborations~\cite{Sin2B_ccK}. These studies measure the asymmetry
amplitude \sinbb, where
$\beta= -\mathrm{arg}(V_{cd}V_{cb}^*/V_{td}V_{tb}^*)$ is a phase
in the Cabibbo-Kobayashi-Maskawa (CKM) quark-mixing
matrix~\cite{CKM}, {\boldmath $V$}.
The \CP violating phase $2\beta$, inferred from \sinbb, has a two-fold
ambiguity, $2\beta$ and $\pi-2\beta$ (four-fold
ambiguity in $\beta$). 
This ambiguity can be resolved by studying decay modes that involve multi-body
final states $\Bz\ra J/\psi \KS\piz$~\cite{ref:C2B},
$D[\KS\pip\pim]\hz$~\cite{Krokovny:2006sv},
$\Dstarp\Dstarm\KS$~\cite{Aubert:2006fh}  or 
$\Kp\Km\Kz$~\cite{ref:KKK0}, where the knowledge of the variation of the
strong phase differences as a function of phase space
allows one also to measure \cosbb.

In this Letter, we present a study of \CP asymmetry in 
$\Bz\ra \DDstar\hz$~\cite{Dsymbol} decays
 with a time-dependent Dalitz plot analysis of
$\D\ra\KS\pip\pim$~\cite{Bondar:2005gk}, where \hz is a \piz, $\eta$, $\etap$,
or $\omega$ meson.
The $\Bz\ra \DDstar\hz$ decay is
dominated by a color-suppressed $\bbar\ra\cbar\u\dbar$ tree amplitude.
The diagram $\bbar\ra\ubar\c\dbar$, which involves a different weak
phase, is suppressed by $V_{ub}V^*_{cd}/V_{cb}V^*_{ud}\simeq 0.02$.
Neglecting the suppressed amplitude,
we factorize the decay amplitude of the chain $\Bz\ra\Dzb\hz\ra
[\KS\pip\pim]\hz$ into $\Af = \A_B\ADbar$ and similarly for \Bzb 
into $\Abarfbar = \A_{\Bbar}\AD$.
The \Dz and \Dzb decay amplitudes are functions of the Dalitz plot variables
$\AD= f(\msp,\msm)$ and $\ADbar= \fbar(\msp,\msm)= f(\msm,\msp)$, where
$\mspm \equiv \msqKspm$.
In the $\Upsilon(4S)\ra\Bz\Bzb$ system,
the rate of a neutral $B$ meson decaying
at proper decay time \trec, the other $B$ (\Btag) at \ttag, and the \D
decaying at a point on the Dalitz plot, is proportional to
\begin{align}
\label{eq:1}
&\frac{e^{-\Gamma\dt}}{2} |\A_B|^2 \cdot
\Big[ (|\ADbar|^2+ |\lambda|^2 |\AD|^2) \notag \\
&\;\;\;\;\;\; \mp  (|\ADbar|^2- |\lambda|^2 |\AD|^2) \cos(\dm\dt)  \\
&\;\;\;\;\;\; \pm 2|\lambda| \xi_{\hz} (-1)^L
\mathrm{Im}(e^{-2i\beta}\AD\ADbar^{*})\sin(\dm\dt)\Big]\,,  \notag
\end{align}
where the upper
(lower) sign is for events with \Btag decaying as a \Bz (\Bzb),
$\dt=\trec-\ttag$, $\Gamma$ is the decay rate of the neutral $B$
meson, $\lambda= e^{-2i\beta} (\A_{\Bb} /\A_B) $, \dm is the \Bz-\Bzb mixing
frequency, $\xi_{\hz}$ is the \CP eigenvalue of \hz, and $(-1)^L$ is the
orbital angular momentum factor.
Here we have assumed \CP-conservation in mixing and
neglected decay width differences.
 For $\D\hz$ modes, $\xi_{\hz}(-1)^L = -1$. For 
$\Dstar[\D\piz]\hz$ ($\hz\neq\omega$) modes $\xi_{\hz}(-1)^L = +1$
including factors from \Dstar decay~\cite{Bondar2}.
In the last term of expression~\ref{eq:1} we can rewrite
\begin{align}
\mathrm{Im}(e^{-2i\beta}\AD\ADbar^{*}) & = \mathrm{Im}(\AD\ADbar^*)\cosbb
\notag \\
& - \mathrm{Re}(\AD\ADbar^*)\sinbb\,,
\end{align}
and treat \cosbb and \sinbb as independent parameters.


We fully reconstruct $\Bz\ra \DDstar\hz$ candidates from a data sample of
$(383\pm 4)\times 10^{6}$ \FourS decays into
$\BB$ pairs collected with the \babar\
detector at the asymmetric-energy $\epem$ PEP-II collider.
The \babar\
detector is described in detail elsewhere~\cite{babar}.
The decay modes used are $\D\piz$, $\D\eta$, $\D\etap$, $\D\omega$,
$\Dstar\piz$, and $\Dstar\eta$, with 
$\Dstar\ra\D\piz$, $\D\ra\KS\pip\pim$, $\KS\ra\pip\pim$, $\piz\ra\gaga$,
$\eta\ra\gaga,\,\piz\pip\pim$, $\etap\ra\eta\,\pip\pim$, and 
$\omega\ra \piz\pip\pim$.

Charged tracks are considered to be pions. 
The \KS candidate is reconstructed from $\pip\pim$ pairs,
whose $\chi^2$ probability of forming a common vertex
is greater than 0.1\%, with invariant mass within $10$\mevcc of the
nominal \KS mass~\cite{nominal}. 
The distance between the \KS decay vertex and the primary interaction point
projected on the $x$-$y$ plane (perpendicular to the beam axis) is required to
be greater than three times its measurement uncertainty. The angle $\theta_K$ 
between the \KS momentum and the line connecting the production and decay
vertices of the \KS on the $x$-$y$ plane is required to satisfy
$\cos\theta_K>0.992$.

An energy cluster in the electromagnetic calorimeter, isolated
from any charged tracks and with the expected lateral shower shape for
photons, is considered a photon candidate. 
A pair of photons forms a $\piz\ra\gaga$ ($\eta\ra\gaga$)
candidate if both photon energies 
exceed 30 (100)\mev and the invariant mass of the pair is between 100 and
160\mevcc (508 and 588\mevcc). If the $\eta$ is paired with a $\Dstar$, the
invariant mass window is tightened to $515 < m_{\gaga} < 581\mevcc$.
The $\eta\ra\gaga$ candidate is rejected if either
photon, when combined with any other photon in the event, 
has an invariant mass within 6\mevcc of the nominal \piz mass. 
We perform a kinematic fit to the photon pair with its invariant mass
constrained at the nominal \piz or $\eta$ mass and reject candidates with a
fit probability less than 0.1\%.

The $\eta / \omega \ra\piz\pip\pim$, $\etap\ra\eta\,\pip\pim$, and
$\D\ra\KS\pip\pim$ candidates are 
formed by combining a \piz, $\eta$, or \KS with two charged pions.
The $\chi^2$ probability of the decay products coming from a common vertex for
\hz (\D) is required to be greater than 0.1\% (1\%).
The momentum of the \piz and the $\eta$ candidates used in $\omega$
and \etap reconstruction must be greater than 200\mevc.
The invariant
masses of the $\eta$, \etap, and $\omega$ candidates are required to be within
10, 8 and 18\mevcc of their respective nominal masses, which correspond to
approximately twice the RMS of the signal distributions.
We retain \D candidates within 60\mevcc of the nominal \Dz mass, 
approximately 10 times its mass resolution, to include sufficient data in the
sideband. 
A kinematic fit is performed on the \D candidate to constrain its mass to the
nominal \Dz mass.
A $\Dstar\ra\D\piz$ candidate is accepted
if the invariant mass difference 
between \Dstar and \D candidates is within 3\mevcc of the nominal mass
difference.

The signal is characterized by the kinematic variables
$\mES\equiv \sqrt{(s/2 + {\mathbf p}_0 \cdot {\mathbf p}_B)^2 / E_0^2 -
{\mathbf p}_B^2}$, and $\DE \equiv E^*_{\B} - E_{\rm beam}^*$,
where the asterisk denotes the quantities evaluated in the center-of-mass (c.m.)
frame, the 
subscripts $0$, beam and \B denote the \epem system, the beam and the \B
candidate, respectively, and $\sqrt{s}$ is the c.m.~energy.
For signal events, \mES
peaks near the $\Bz$ mass with a resolution of about 3\mevcc, and
\DE peaks near zero, with a resolution that varies by mode.
We require $\mES>5.23\gevcc$
and select events with $|\DE|< 80\mev$ for modes with $\piz,\eta\ra\gaga$, and 
$|\DE|< 40\mev$ for modes with $\eta,\omega\ra\piz\pip\pim$, or $\etap\ra\eta\,\pip\pim$.

The proper decay time difference \dt is determined from the measured distance
between the two \B decay vertices projected onto the boost axis
and the boost ($\beta\gamma= 0.56$) of the c.m. system.
The reconstructed $|\dt|$ and its uncertainty \sigmadt are required
to satisfy $|\dt|<15$~ps and $\sigmadt<2.5$~ps.
The flavor of \Btag is identified from particles that do not belong
to the reconstructed \B meson using a neural network based flavor-tagging
algorithm~\cite{tag}.

The main background is from the continuum $\epem\ra\qqbar$ ($q = u, d, s, c$). 
We use a Fisher discriminant (${\cal F}$) to separate the more
isotropic \BB events from more jet-like $\qqbar$ events~\cite{d0h0s2b}. The
requirement on ${\cal F}$ is optimized with simulation.
 Another major background for the
$\Dstar\piz$ mode comes from color-allowed $\Bm\ra \Dz\rho^-
(\rho^-\ra\piz\pim)$ decays, which mimic signal if the \pim is missed from
reconstruction while a random \piz is included. 
We veto the \Bz candidate if the combination of another charged pion in the
event with the \D and the \piz in the \Bz candidate is consistent with a
charged \B decay.
In total we select 4450
events, of which 2843 events have useful tagging information (tagged).

\begin{figure}[tb]
\begin{center}
\includegraphics[width=0.235\textwidth]{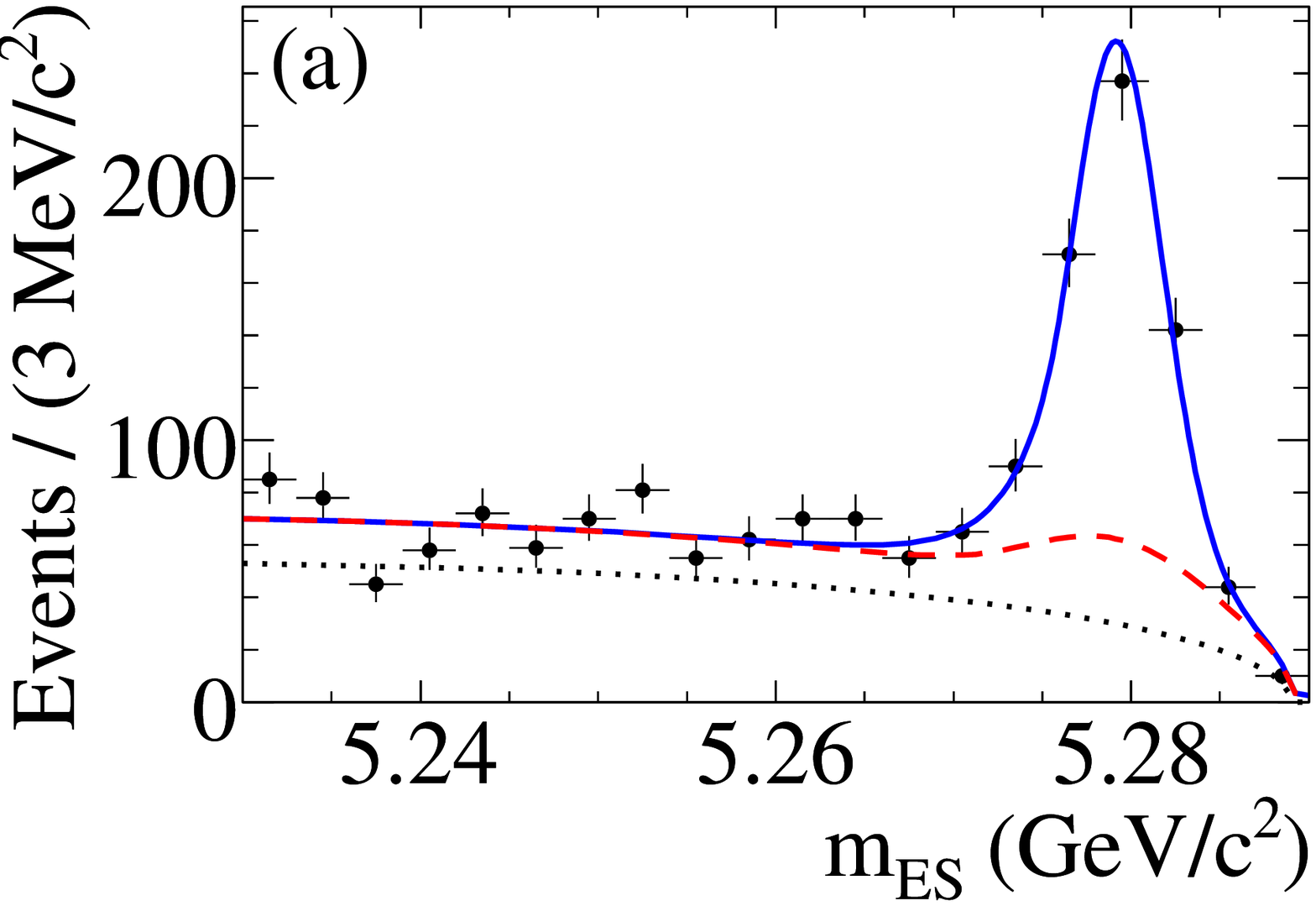}
\includegraphics[width=0.235\textwidth]{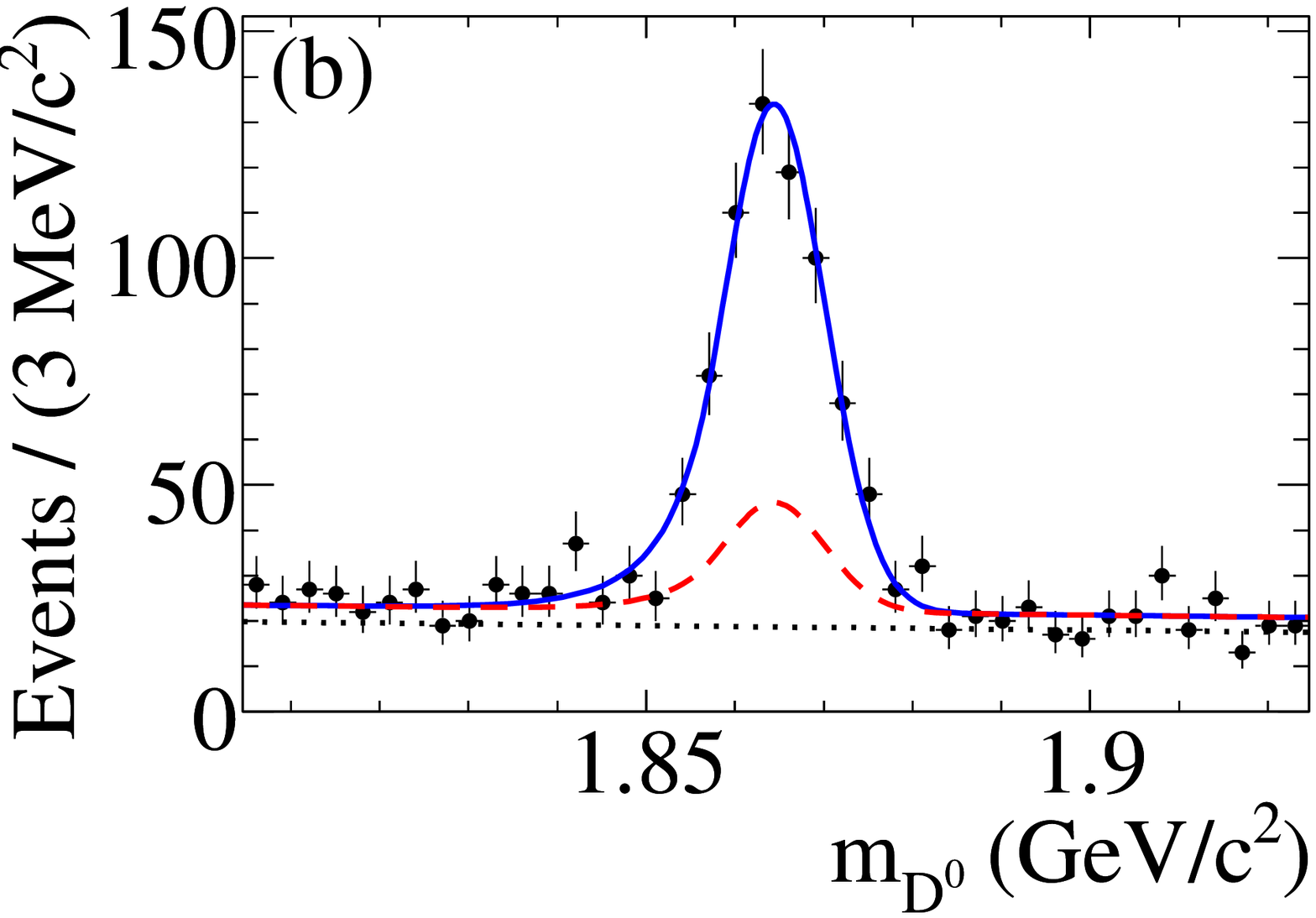}
\end{center}
\caption{
Distributions of (a) \mES [$|\mD-\mDz^{\mathrm{PDG}}|<14\mevcc$], and
(b) \D mass [$\mES>5.27\gevcc$].
Dashed (dotted) lines represent the total (non-peaking) background. 
}
\label{fig:mesmd0}
\end{figure}

The signal and background yields are determined by a fit to the $(\mES,\mD)$
distributions using a two-dimensional probability density function (PDF),
where \mD denotes the $\KS\pip\pim$ invariant mass. We
divide the sample into four categories to take into account different
background levels: (1) $\D\piz$,  (2) $\D\eta$ and $\D\etap$ (3)
$\D\omega$, 
and (4) $\Dstar\hz$. The PDF has five components: (a) signal, and backgrounds
that peak in (b) both \mES and \mD, (c) \mES but not \mD, (d) \mD but not
\mES, and (e) neither distribution.
Both peaks are modeled
by a Crystal Ball line shape~\cite{crystal_ball}. The non-peaking component is
modeled by a straight line in \mD and a threshold function~\cite{argus} in
\mES. We fit the four categories of events simultaneously, allowing the \mES
peak shape to be different but letting them share the \mD shape and \mES
background parameters. We first determine the amount of the peaking component
(b) from simulated events and then fit to data allowing all other components
to vary. We obtain $463\pm39$ signal events ($335\pm32$ tagged). The
contribution from each mode is shown in Table~\ref{tab:table1}. The \mES and \mD
distributions are shown in Fig.~\ref{fig:mesmd0}.

The $\Dz\ra\KS\pip\pim$ Dalitz plot has been studied in
detail~\cite{belle_gamma,babar_gamma}. We use the isobar
formalism described in~\cite{isobarCLEO} to express \AD as a sum of two-body
decay matrix elements ($\A_r$) and a non-resonant (NR)
contribution,
\begin{equation}
\AD = a_{\mathrm{NR}} e^{i \phi_{\mathrm{NR}}}+
 \sum_r a_r e^{i\phi_r} \A_r(\msp,\msm)\,.
\end{equation} 
The function $\A_r(\msp,\msm)$ is the Lorentz-invariant expression for the
matrix element of a \Dz decaying into $\KS\pip\pim$ through an intermediate
resonance $r$, parameterized as a function of the position on the Dalitz
plot. The resonances in the model include $\Kstar(892)$, $\Kstar_0(1430)$,
$\Kstar_2(1430)$, $\Kstar(1410)$ and $\Kstar(1680)$ for both $\KS\pip$ and
$\KS\pim$, and $\rho(770)$, $\omega(782)$, $f_0(980)$, $f_0(1370)$,
$f_2(1270)$, $\rho(1450)$, and two scalar terms $\sigma$ and $\sigma^\prime$ in
the $\pip\pim$ system. 
Details of the Dalitz model and the parameters (determined from data)
can be found in~\cite{babar_gamma}.


To perform the time-dependent Dalitz plot analysis, we expand the PDF to
include \dt and Dalitz plot dependence. The signal component is proportional to
expression~\ref{eq:1}, modified to account for the probability of mis-identifying
the \Btag flavor (mistag), and is convolved with a sum of three Gaussian
distributions~\cite{babar_s2b_prd}.
The mistag parameters and the
resolution function are determined from a large data control sample of
$\Bz\ra D^{(*)-}h^{+}$ decays, where $h^{+}$ is a $\pip$, $\rho^+$, or
$a_1^{+}$ meson. Each of the background components consists of a product
of \dt and $(\msp,\msm)$ PDFs. The components that peak in \mD use
$\AD(\msp,\msm)$ as their Dalitz model. The model for components that are
flat in \mD is an incoherent sum of a phase space contribution and
several resonances. The choice of resonances and their relative contributions
are determined empirically from events outside the \mD peak.
The \dt model for components 
that peak in \mES is a simple exponential decay convolved with the resolution
function used in the signal component. For the non-peaking background, we
use a zero-lifetime component convolved with a 
double-Gaussian resolution function for events with a real \D because they
are dominated by $\ccbar$ events, and we add
an exponential decay component for events without a real \D to account
for $B$ background.

\begin{figure}
\begin{center}
\includegraphics[width=0.48\textwidth]{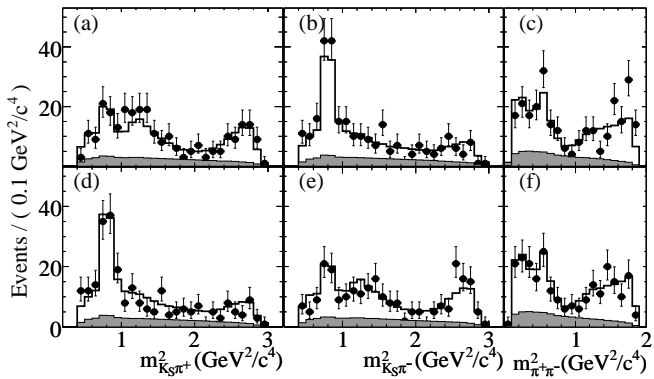}
\end{center}
\caption{Dalitz plot projections for (a,b,c) \Bz-tagged events and (d,e,f)
\Bzb-tagged events. Points are data, open histograms are PDF projections, and
shaded histograms are background contributions.}
\label{fig:Dalitz}
\end{figure}

\begin{figure}
\begin{center}
\includegraphics[width=0.48\textwidth]{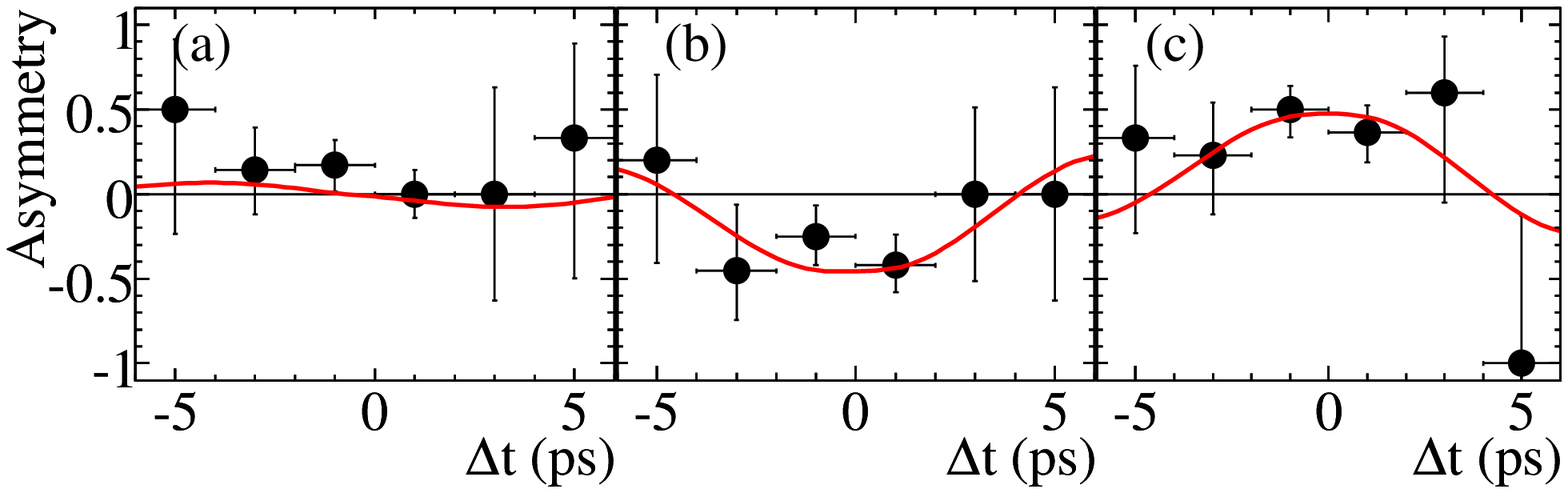}
\end{center}
\caption{Time-dependent asymmetries for (a) $D\ra\KS\rho$ region 
($|m_{\pip\pim}-770|<150$), where the opposite \CP
asymmetry in $\Dstar\hz$ has been taken into account,
(b) $D\ra \Kstarp\pim$
region, and (c) $D\ra \Kstarm\pip$ region ($|m_{\KS\pi}-892|<50$). Units are \mevcc. Curves are projections of the PDF.} 
\label{fig:asym}
\end{figure}

We fit the \mES, \mD, and \dt distributions, with \mES and \mD shapes and
background fractions fixed by the previous fit for event yields, 
to determine the \dt parameters for backgrounds. We then perform the final fit
adding Dalitz plot variables to determine \cosbb, \sinbb and \abslambda.
Table~\ref{tab:table1} shows the nominal fit result (All) and the results of a
fit 
allowing \cosbb and \sinbb to be different among the four types of events.
The correlations
are $\rho(\cosbb,\sinbb)=2\%$, $\rho(\abslambda,\cosbb)=2\%$, and 
$\rho(\abslambda,\sinbb)=-2\%$.
The Dalitz plot projections are shown in Fig.~\ref{fig:Dalitz}. 
Figure~\ref{fig:asym} shows the time-dependent asymmetries 
$(N_+-N_-)/(N_++N_-)$, where $N_+(N_-)$ is the number of $\Bz(\Bzb)$ tagged
events, 
for events in various Dalitz plot regions. Events in the $\D\ra\KS\rho$ region
are dominated by a single \CP eigenstate, thus the asymmetry is proportional to
$\sinbb\sin(\dm\dt)$. Events near $\D\ra\Kstarpm \pi^\mp$ are dominated by
decays to a definite flavor, and therefore exhibit a $\cos(\dm\dt)$ behavior.

\begin{table}[htb]
\caption{\label{tab:table1} Tagged event yields $N_{\mathrm{tag}}$ and
  fit results. Errors are statistical.} 
\begin{center}
\begin{tabular*}{0.475\textwidth}{@{\extracolsep{\fill}}lr@{\,$\pm$\!\!\!}lr@{\,$\pm$\!\!\!}lr@{\,$\pm$\!\!\!}lc}
\hline\hline
Mode & \multicolumn{2}{c}{$N_{\mathrm{tag}}$} & \multicolumn{2}{c}{\cosbb} & \multicolumn{2}{c}{\sinbb} & \abslambda \\
\hline
$\D\piz$ & $143$ & $19$ & $0.78$ & $0.92$ & $0.70$ & $0.52$ & \multirow{4}{*}{1.0 (fixed)}\\
$\D\eta/\etap$ & $60$ & $12$ & $1.20$ & $1.19$ & $-1.17$ & $1.00$
& \\
$\D\omega$ & $76$ & $12$ & $0.43$ & $0.87$ & $-0.48$ & $0.74$ & \\
$\Dstar\hz$ & $56$ & $12$ & $-0.56$ & $1.07$ & $0.78$ & $0.87$ & \\
\hline
All & $335$ & $32 $ &  $0.42$ & $0.49$ & $0.29$ & $0.34$  & $1.01\pm0.08$ \\
\hline\hline
\end{tabular*}
\end{center}
\end{table}

The dominant systematic uncertainty is the Dalitz plot model dependence. The
Dalitz model includes scalar terms $\sigma$ and $\sigma^\prime$, which
are not well established, in order to achieve a good
quality fit~\cite{babar_gamma}.
We study the
effect of these two scalars by simulating a number of datasets, each of which
is 50 times the size of the data, according to the PDF, and repeat the final fit
using both the nominal PDF and the PDF without the two scalars. We compare the
results between the two fits in each dataset and conservatively take the
quadratic sum of the mean and RMS of the differences as the systematic
uncertainty: $\sigma(\cosbb)= 0.13$, $\sigma(\sinbb)= 0.05$, and
$\sigma(\abslambda)< 0.01$.
Many parameters are pre-determined in fits to control samples and to data
without the Dalitz variables. We randomize them according to a Gaussian
distribution whose width equals one standard deviation of each parameter,
taking correlations into account, and repeat 
the final fit. The width of the distribution is taken as the systematic
uncertainty: 
$\sigma(\cosbb)= 0.06$, $\sigma(\sinbb)= 0.02$ from
Dalitz model parameters;
$\sigma(\cosbb)= 0.05$, $\sigma(\sinbb)= 0.02$ from
\mD and \mES shape parameters;
$\sigma(\cosbb)\lesssim 0.01$, $\sigma(\sinbb)\lesssim 0.01$ from
background \dt parameters, tagging parameters, or signal \dt resolution
function.
We also vary the peaking background fractions by the statistical uncertainty
found in simulation and
find the variations are $\sigma(\cosbb)= 0.02$ and $\sigma(\sinbb)= 0.01$.
Other sources of uncertainty such as \Bz-\Bzb mixing frequency, \B lifetimes,
background 
Dalitz model and reconstruction efficiency variation over the Dalitz plot are
negligible. 
In all cases, the uncertainty on \abslambda is less than 0.01.
The only significant uncertainty on \abslambda ($\sim 0.02$) is from
the interference between the CKM-suppressed $\bbar\ra \ubar c
\dbar $ and CKM-favored $b \ra c \ubar d$ amplitudes in some \Btag final
states~\cite{tagsideint}. This effect is studied with simulation.
Summing over all contributions in quadrature, we obtain total experimental
systematic uncertainties 
$\sigma(\cosbb)= 0.09$, $\sigma(\sinbb)= 0.03$, and
$\sigma(\abslambda)= 0.02$.

To resolve the ambiguity in $2\beta$, we generate two sets of toy simulation
samples, one with $\cosbb= \sqrt{1-S_0^2}\equiv C_0$, and the other with
$\cosbb= -C_0$, 
where $S_0= 0.678$, the world average of \sinbb~\cite{HFAG}, and fit each
sample while fixing $\sinbb= S_0$ and $\abslambda=1$. For data, this
configuration results in $\cosbb= 0.43\pm 0.47$.
We then use double-Gaussian functions, $h_\pm(x)$ for $\pm C_0$ hypotheses, to
model the probability density of the resulting \cosbb distributions, smeared
by the 
experimental systematic uncertainty and the uncertainty of $C_0$.
The confidence level (C.L.) of preferring $\cosbb= +C_0$ over $-C_0$ is
defined as $h_+(x)/[h_+(x)+h_-(x)]$ if $\cosbb=x$ is observed in data. 
Considering the Dalitz model dependence for \cosbb (0.13), we use $x$ between
$0.43\pm 0.13$ and find the smallest C.L.$=86\%$ at $x=0.43-0.13$.

In conclusion, we have studied the $\Bz\ra \DDstar\hz$ decays using a
time-dependent Dalitz plot analysis of $\D\ra\KS\pip\pim$. We obtain 
$\cosbb =  0.42 \pm 0.49(\mathrm{stat.}) \pm 0.09(\mathrm{syst.}) \pm 0.13(\mathrm{Dalitz}) $, 
$\sinbb =  0.29 \pm 0.34(\mathrm{stat.}) \pm 0.03(\mathrm{syst.})  \pm 0.05(\mathrm{Dalitz}) $, and 
$\abslambda = 1.01 \pm 0.08(\mathrm{stat.}) \pm 0.02(\mathrm{syst.}) $. 
Using the world average $\sinbb=0.678\pm 0.026$ and
$\abslambda=1$, $\cosbb>0$ is 
preferred over $\cosbb<0$ at 86\% C.L.

We are grateful for the excellent luminosity and machine conditions
provided by our \pep2\ colleagues, 
and for the substantial dedicated effort from
the computing organizations that support \babar.
The collaborating institutions wish to thank 
SLAC for its support and kind hospitality. 
This work is supported by
DOE
and NSF (USA),
NSERC (Canada),
CEA and
CNRS-IN2P3
(France),
BMBF and DFG
(Germany),
INFN (Italy),
FOM (The Netherlands),
NFR (Norway),
MIST (Russia),
MEC (Spain), and
STFC (United Kingdom). 
Individuals have received support from the
Marie Curie EIF (European Union) and
the A.~P.~Sloan Foundation.


\end{document}